\documentclass[12pt]{article}
\usepackage[latin1]{inputenc}

\usepackage{amsmath}
\usepackage{amsfonts}
\usepackage{amssymb}
\usepackage{graphicx}
\usepackage{geometry}
\usepackage{amssymb,epsfig,subfigure}
\usepackage{hyperref}

\def\simge{
    \mathrel{\rlap{\raise 0.511ex 
        \hbox{$>$}}{\lower 0.511ex \hbox{$\sim$}}}}

\def\simle{
    \mathrel{\rlap{\raise 0.511ex 
        \hbox{$<$}}{\lower 0.511ex \hbox{$\sim$}}}}

\makeatletter
\renewcommand\section{\@startsection {section}{1}{\z@}%
                                 {-3.5ex \@plus -1ex \@minus -.2ex}
                                   {2.3ex \@plus.2ex}%
                                   {\normalfont\large\bfseries}}
\renewcommand\subsection{\@startsection{subsection}{2}{\z@}%
                                   {-3.25ex\@plus -1ex \@minus -.2ex}%
                                     {1.5ex \@plus .2ex}%
                                     {\normalfont\bfseries}}
\renewcommand\subsubsection{\@startsection{subsubsection}{3}{\z@}%
                                   {-3.25ex\@plus -1ex \@minus -.2ex}%
                                     {1.5ex \@plus .2ex}%
                                     {\normalfont\itshape}}
\makeatother

\def\pplogo{\vbox{\kern-\headheight\kern -29pt
\halign{##&##\hfil\cr&{\ppnumber}\cr\rule{0pt}{2.5ex}&\ppdate\cr}}}
\makeatletter
\def\ps@firstpage{\ps@empty \def\@oddhead{\hss\pplogo}%
  \let\@evenhead\@oddhead 
}
\def\maketitle{\par
 \begingroup
 \def\thefootnote{\fnsymbol{footnote}}
 \def\@makefnmark{\hbox{$^{\@thefnmark}$\hss}}
 \if@twocolumn
 \twocolumn[\@maketitle]
 \else \newpage
 \global\@topnum\z@ \@maketitle \fi\thispagestyle{firstpage}\@thanks
 \endgroup
 \setcounter{footnote}{0}
 \let\maketitle\relax
 \let\@maketitle\relax
 \gdef\@thanks{}\gdef\@author{}\gdef\@title{}\let\thanks\relax}
\makeatother

\numberwithin{equation}{section}
\newcommand{\tab}{\hspace*{2em}}

\newcommand{\be}{\begin{equation}}
\newcommand{\bea}{\begin{eqnarray}}
\newcommand{\ee}{\end{equation}}
\newcommand{\eea}{\end{eqnarray}}
\newcommand\beq{\begin{equation}}
\newcommand\eeq{\end{equation}}

\newcommand{\mc}{\mathcal}

\newcommand{\tE}{{\tilde E}}
\newcommand{\tp}{{\tilde p}}
\newcommand{\tq}{{\tilde q}}

\begin{document}

\setcounter{page}0
\def\ppnumber{\vbox{\baselineskip14pt
}}
\def\ppdate{\footnotesize{SLAC-PUB-15119, SU-ITP-12/18}} \date{}

\author{Ning Bao, Xi Dong, Sarah Harrison, and Eva Silverstein\\
[7mm]
{\normalsize \it Stanford Institute for Theoretical Physics }\\
{\normalsize  \it Department of Physics and SLAC, }\\
{\normalsize \it Stanford, CA 94305, USA}\\
[3mm]}

\bigskip
\title{\bf  The Benefits of Stress:\\
Resolution of the Lifshitz Singularity
\vskip 0.5cm}
\maketitle

\begin{abstract}

Through the AdS/CFT correspondence, Lifshitz spacetimes describe field theories with dynamical scaling ($z \neq 1$).  Although curvature invariants are small, the Lifshitz metric exhibits a null singularity in the IR with a large tidal force that excites string oscillator modes.  However, Lifshitz is not a vacuum solution of the Einstein equations -- the metric is supported by nontrivial matter content which must be taken into account in analyzing the propagation of test objects.  In this paper, we consider the interaction of a string with a D0-brane density in the IR which supports a class of UV-complete $z=2$ Lifshitz constructions.  We show that string/D-brane scattering in the Regge limit slows the string significantly, preventing divergent mode production and resolving the would-be singularity in string propagation.

\end{abstract}
\bigskip
\newpage

\tableofcontents

\vskip 1cm

\section{Introduction}
The holographic duality between gravity and field theory \cite{Maldacena}\ has proved effective in addressing very interesting questions in finite density field theory inspired from condensed matter physics  \cite{Rev1, Rev2, Rev3}.  
At finite density, since the Lorentz symmetry is broken, one might expect to find a variety of quantum critical theories respecting a Lifshitz scaling symmetry 
\be\label{scaling}
t\rightarrow \lambda^z t, \tab x\rightarrow \lambda x.
\ee
with nontrivial $z$.
Aside from weakly interacting systems, such behavior is not easy to establish either theoretically or experimentally.  Hence a holographic analysis is well worth developing \cite{KLM}, as it provides a novel handle on some strongly interacting field theories.  The problem of modeling non-Fermi liquid (NFL) behavior also strongly motivates the development of a better understanding of Lifshitz systems, since they provide simple mechanisms for  NFL transport such as $T$-linear resistivity  in certain classes of field theories \cite{HPST, CMz}.    

This scaling symmetry is realized holographically in what is known as the Lifshitz metric \cite{KLM}
\be\label{intrometric}
ds^2=L^2\left(-r^{2z}dt^2+\frac{dr^2}{r^2}+r^2dx_i^2\right)
\ee
supported by stress-energy from additional fields.\footnote{To describe heating up the system, black hole solutions with Lifshitz asymptotics were obtained in \cite{Mann, bert, McG, Thorl, Amado:2011nd}.  The extension of the holographic dictionary to this case was analyzed in \cite{Ross}.  Various methods of constructing this metric as solutions of general relativity with specific matter content have been found \cite{Taylor, Gubser, HPST, Hartnoll}. Realizations of Lifshitz in string theory and supergravity have been developed in \cite{HPST, Balas, Donos, Gregory, Cassani:2011sv}.}  In this paper, the sources supporting the Lifshitz metric will play a crucial role, along with the degrees of freedom arising in its UV completion.

The Lifshitz metric has a divergent tidal force as $r\rightarrow 0$ (for $z\neq 1,\infty$) \cite{KLM, CM}\ along a null singularity, as we will review briefly below.  In an interesting variant studied recently \cite{Resolving}\ -- a
 system with a running coupling dual to a radially varying string coupling on a Lifshitz metric -- a similar would-be singular region is strongly coupled on the gravity side, and the bottom-up solution ultimately flows to $z=\infty$.  It may be possible to find other  examples of the kind suggested in \cite{Balasubramanian:2011ua}, exhibiting a running coupling and confinement in the infrared instead of a persistent Lifshitz scaling (but see \cite{Copsey:2011ek}).  It is also possible to deform away from the singular solution by moving the system out onto its approximate moduli space; this provides a method of forming the system from nonsingular initial data by sending the scalar fields back to the origin of moduli space.\footnote{Some interesting aspects of the relevant brane propagation in Lifshitz geometries were studied in \cite{Alishahiha}.}  For many applications, a wide window of scales with Lifshitz scaling may suffice, with instabilities to superconductivity or other phases taking over in the deep infrared.        

However, the question persists as to the deep-IR fate of pure Lifshitz systems (without running couplings); as in the Lorentz invariant case fixed points play a special role in organizing quantum field theories.  To understand this, one must address the behavior of excitations in the theory which are sensitive to the tidal forces. This was considered in \cite{Horoway}\ for strings propagating in the Lifshitz metric, where divergent mode production was obtained for strings going through the singularity into an additional region introduced to control the analysis of Bogoliubov coefficients.  As an initial step in the present paper, we work within the original Lifshitz geometry and show how the tidal force introduces a strongly time dependent mass term on the string worldsheet, confirming that a purely gravitationally coupled string would indeed experience singular mode production.  The interpretation proposed in \cite{Horoway}\ was that Lifshitz solutions are unstable in string theory.          

But the string also interacts with the stress-energy sources which support the Lifshitz metric, and this may affect its propagation and mode production.  
In this paper, we will focus on a specific UV theory discussed in \cite{HPST}, where a density of D0-branes sources a Lifshitz metric in simple Freund-Rubin compactifications of type IIA string theory.  The interaction with D0-branes slows the probe string sufficiently to avoid the infinite mode production obtained in \cite{Horoway}\ at the level of the metric.  Other objects such as D-branes are also affected by the D0-brane density, getting trapped and slowed through off-diagonal string production.
It will be interesting to apply this lesson to other UV completions of Lifshitz such as \cite{Balas, Donos, Gregory, Cassani:2011sv}, where the geometry is supported by flux densities, and to other similar geometries such as those in \cite{Edgar}\ which 
exhibit diverging mode production at the level of the metric.

\section{The Lifshitz singularity at the level of the metric}\label{Lif}

The Lifshitz metric is
\be\label{metric}
ds^2=L^2\left(-r^{2z}dt^2+\frac{dr^2}{r^2}+r^2dx_i^2\right) \,,
\ee
where $i$ runs over the spatial directions of the dual field theory.  
If we ignore the stress-energy sources with which they interact, and also for now ignore internal excitations introduced by tidal forces, then massive particles follow geodesics derived from the effective action 
\beq\label{BI}
{\cal S}=-mL\int d\tau\sqrt{r^{2z}\dot t^2-\frac{\dot r^2}{r^2}-r^2{\dot x_i^2}}\,,
\eeq
where dots mean derivatives with respect to the proper time $\tau$.
If we consider radial motion, these
propagate along a trajectory
\beq\label{mgeodesics}
\dot{t}=\frac{E}{mL r^{2z}} \,,\quad
\dot{r}=\frac{E}{mLr^{z-1}}\sqrt{1-\frac{m^2 r^{2z}}{E^2}} \,,
\eeq
in terms of the conserved energy \cite{CM}
\be\label{Econserved}
E=-\frac{m g_{00}}{L}\dot t=\frac{\gamma m \sqrt{-g_{00}}}{L}=\gamma m r^z \,.
\ee

These trajectories imply divergent tidal forces as $r\to 0$ for objects extended in the $x^i$ directions \cite{KLM, CM, Horoway}.
Heuristically, this arises because
neighboring particles propagate radially on trajectories of constant coordinate separation $\Delta x$,
leading to a shrinking proper separation $rL\Delta x$ with divergent relative acceleration for $z\ne 1$:
\beq\label{heuristic}
\frac{d^2 r}{d\tau^2} \sim \frac{E^2(1-z)}{m^2L^2 r^{2z-1}} ~~~~ {\rm as}~~r\to 0.
\eeq  
The covariant quantity capturing this effect is the geodesic deviation 
\beq\label{geodev}
T^\mu \nabla_\mu\left(T^\nu \nabla_\nu\hat{x}\right)\sim \frac{E^2(1-z)}{m^2L^2r^{2z}} ~~~~ {\rm as}~~r\to 0 \,,
\eeq 
given in terms of the tangent vector to the trajectory $T=(\dot t, \dot r, 0,0)$, and a vector $\hat x=(0,0,1,0)$
transverse to it.  A similar result holds for massless trajectories.  An extended object such as a string is sensitive to the tidal force, and it was argued in \cite{Horoway}\ that this induces rampant mode production, raising the question of whether field theories dual to Lifshitz geometries such as those in \cite{HPST}\ are truly scale invariant in the deep IR.       

We can exhibit the effect of (\ref{geodev}) on the string oscillator spectrum in the following simple way.
The bosonic part of the worldsheet action is
\beq\label{Sws}
{\cal S}_{ws}\sim \frac{1}{\alpha'}\int d\sigma d\tau G_{MN}(X)\partial X^M \partial X^N  \,.
\eeq
To see the effect of the tidal force on mode production, let us evaluate this in the trajectory (\ref{mgeodesics}).  This will also enable us to check whether the induced excitation of the string changes its propagation significantly relative to (\ref{mgeodesics});
the energy $E$ is conserved, and exciting the string will slow it down to some extent. 
In the background trajectory (\ref{mgeodesics}), the action for the embedding coordinates $X$ in the $x_i$ directions is
\bea\label{Swstraj}
{\cal S}_{ws,\perp} &\sim& \frac{1}{\alpha'}\int d\sigma d\tau r(\tau)^2\left(\dot X^2-X'^2\right) \nonumber\\
&=& \int d\sigma d\tau \left\{ \left(\frac{d\tilde X}{d\tau}\right)^2+\frac{1}{r}\frac{d^2 r}{d\tau^2}\tilde X^2-\left(\frac{d\tilde X}{d\sigma}\right)^2\right\}
\eea
where in the last expression we changed variables to $\tilde X=r(\tau)X/\sqrt{\alpha'}$ and did one integration by parts.
This expression exhibits a strongly time-dependent worldsheet mass squared $\tilde m^2=-\ddot r/r$ which is precisely equal to the tidal force (\ref{geodev}) (as can be seen by comparing (\ref{heuristic}) with (\ref{geodev})).  Including the gradient term, we have a $\tau$-dependent frequency $\tilde\omega$ for the worldsheet fields $\tilde X$:
\beq
\tilde\omega^2 =\tilde m^2 + \frac{\tilde n^2}{\ell^2} =-\frac{1}{r} \frac{d^2 r}{d\tau^2} + \frac{\tilde n^2}{\ell^2}
\eeq
where $\tilde n$ is the worldsheet mode number and $\ell$ is its length.

To estimate mode production from this, we note that the system will be non-adiabatic when $\dot{\tilde\omega}/\tilde\omega^2\simge 1$:
\beq\label{nonadiabatic}
\frac{\dot{\tilde\omega}}{\tilde\omega^2}=\frac{\frac{1}{2}\frac{d}{d\tau}(\tilde m^2) }{\tilde\omega^3}
\sim\frac{1}{\sqrt{z-1}}\frac{1}{1+\frac{\tilde n^2}{\ell^2\tilde m^2}} \,.
\eeq 
(Note that $\tilde m^2\propto z-1$ so the expression goes to zero for $z=1$.) 
Thus the tidal force induces mode production up to a mode number
\beq\label{osc}
\frac{\tilde n}{\ell}\sim \tilde m\sim \sqrt{z-1}\frac{\gamma}{L} \,.
\eeq
This translates into an increase in the proper string mass $m$ of order
\beq\label{tidalmass}
\Delta m \sim \sqrt{z-1}\frac{\gamma}{L} \,,
\eeq
slowing it down relative to the geodesic trajectory (\ref{mgeodesics}) of a massive particle.

However, as discussed in the introduction, the propagation of objects in a Lifshitz background in general depends on more than the metric; we must include interactions with the stress-energy sources supporting the Lifshitz metric, interactions which are dictated by a UV complete realization such as \cite{HPST}\ or \cite{Balas, Donos}.  

Before performing this stress(-energy) test in the next two sections, let us work out sufficient conditions under which the trajectories would not experience divergent forces.  First, note that if the speed $v$ of the probe is bounded from above by some subluminal value, it takes infinite proper time to reach $r=0$:  on such a trajectory
\beq\label{termvelocity}
d\tau^2 =-L^2 \left(\frac{dr}{r}\right)^2\left(1-\frac{1}{v^2}\right).
\eeq 
Suppose such a trajectory arose through a process of scattering, free fall, further scattering, and so on.
During free fall, the tidal force is calculated as above.  To see what happens, let us plug \eqref{Econserved} into \eqref{geodev} and find
\be\label{finiteforce}
T^\mu \nabla_\mu\left(T^\nu \nabla_\nu\hat{x}\right)\sim \frac{\gamma^2(1-z)}{L^2} \,.
\ee
For finite $\gamma$, we now see that the tidal force no longer diverges as $r\to0$.

In the following sections we will show that the late-time $\gamma$ is bounded for strings in concrete examples of Lifshitz spacetimes, at a value well below the crossover between string perturbation theory and nonperturbative physics.  We will find more generally that divergent mode production does not occur.

\section{Top-down Lifshitz sources and string propagation}\label{topdown}

In this section, let us begin by reviewing a specific top-down construction of Lifshitz in \cite{HPST}, paying special attention to a density of D0-branes that support the solution.  We anticipate that these D0-branes will affect the propagation of strings and other objects in the Lifshitz background.

\subsection{Setup}

As explained in \cite{HPST}, flows from 2+1 dimensional CFTs dual to $AdS_4$ in the UV to $z=2$ Lifshitz in the IR arise in type IIA Freund-Rubin compactifications, sourced by a density of D0-branes below some radial scale along with the radial electric field under which they are charged.  Examples of the underlying 2+1 dimensional CFT include the field theory on the D2-D6 system (dual to type IIA on $AdS_4\times S^6$ with flavor D6-branes wrapping an $S^3\subset S^6$),
and the ABJM theory (dual to type IIA on $AdS_4\times CP^3$ with 6-form and 2-form RR fluxes).

In order to analyze string propagation, we need the scaling of the curvature radius $L$, the string coupling $g_s$, and the D0-brane energy density $\rho_0$ with the discrete quantum numbers in the construction.  To be concrete, let us review this for the D2-D6 case with $N_2$ D2 color branes and $N_6$ D6 flavor branes, but the ABJM case is similar.

The scalings are obtained from the effective action by equating the Einstein--Hilbert term, flux kinetic energy, and D0-brane density up to order 1 factors, since each term contributes at leading order to the solution.  This yields the scalings
\beq\label{DtwoDsixscalings}
\frac{L^4}{\alpha'^2}\sim \frac{N_2}{N_6} \,,\qquad
g_s\sim\frac{N_2^{1/4}}{N_6^{5/4}} \,,\qquad
\alpha'^2\rho_0\sim \frac{L^4}{g_s^2} \,.
\eeq

We can now determine the number density and spacing between the D0-branes.  This requires specifying their distribution in all ten dimensions.  Let us consider two extremes, either putting them at the same point on the internal $S^6$, or  distributing them uniformly in all dimensions.  In the former case, 
recalling that $\rho_0$ is the proper mass density of D0-branes, we find the number density $n_0$ and the average proper distance $\delta$ between neighboring D0-branes to be
\be\label{numden}
\alpha'^{3/2} n_0 \sim \frac{L^4}{g_s\alpha'^2} \,,\qquad
\delta \sim \frac{g_s^{1/3}}{L^{4/3}}\alpha'^{7/6} \,.
\ee
The average distance $\delta$ is parametrically larger than the 4D Planck scale ($\sim \alpha'^2 g_s/L^3$).
If instead we distribute the D0-branes uniformly in all nine spatial dimensions, including on the $S^6$ of radius $L$, we obtain a 10d number density
\beq\label{numdenten}
n_0^{10d}\sim \frac{n_0}{L^6}\sim \frac{1}{g_s L^2\alpha'^{7/2}} \sim \frac{{N_6}^{7/4}}{N_2^{3/4}}\frac{1}{\alpha'^{9/2}}
\eeq
For simplicity, let us work with this configuration in what follows, so as to avoid having to consider different possibilities for the position of the probe string on the $S^6$.
We note here that it is straightforward to obtain a regime ($N_6^{7/3}\ll N_2\ll N_6^5$) in which the D0-branes are separated by more than string scale, consistently with the requirements that $g_s$ and $\alpha'^2/L^4$ both be small.  This is convenient for our analysis of string/D0-brane scattering below.\footnote{In general, it would be interesting to consider various distributions of the D0-branes (including cases in which they are clumped together in larger groups).}    

Next, we note that a string propagating to the horizon will collide with an infinite number of these D0-branes before reaching the horizon, since the proper D0-brane density is constant and the proper distance from any $r$ to the horizon is infinite:
\be
n_0\int_0 L\frac{dr}{r}=\infty \,.
\ee
This does not contradict the finite charge density in the dual field theory, as the charge density is calculated by counting the number of D0-branes in a cylinder that is extended in the $r$ direction and has a constant \textit{coordinate} area.  The proper area becomes small near the horizon and captures very few D0-branes, leading to a finite charge density
\be
n_0\int_0^{r_0} r^2L\frac{dr}{r}=\frac12 n_0Lr_0^2
\ee
in the field theory, where $r_0$ is the radial position below which D0-branes are present.

\subsection{Energy loss and gain}

In this section, we will determine and compare the proper energy lost from scattering with the proper energy gained as the string falls toward $r=0$.

To start with, let us estimate how many times a string will collide with D0-branes while traveling a distance over which gravity takes effect, i.e.~a distance comparable to the curvature radius $L$.  
The cross section $\sigma$ of a string scattering off a D0-brane can be conveniently written as $\sigma=g_s^2{\hat\sigma}$, where we have made explicit the dependence on $g_s$.  The expected number of collisions in a proper distance of order $L$ is then   
\be\label{scatteringnumber}
N_{collisions}|_L\sim  n^{10d}_0 L \sigma \sim \frac{g_s}{L\alpha'^{7/2}}{\hat\sigma}
\ee
where we have used \eqref{numdenten}.

Before moving to a discussion of the cross section $\sigma$, let us determine the proper energy gained by a string freely falling a proper distance $\sim L$ in the Lifshitz geometry.  
In terms of the coordinate $r$, a proper distance $L$ corresponds to traveling from an initial radial position $r$ to a final radial position $r/e$.  
From \eqref{Econserved} we have the proper energy
\beq\label{Eproper}
\tE=\gamma m = \frac{E}{r^z} \,.
\eeq
Here and below we use tilded variables to denote the proper energy $\tE=\gamma m$ and momentum $\tp = \gamma m v$ to avoid confusion with the conserved energy $E$ described above (conjugate to the time coordinate $Lt$) and the conserved momentum (conjugate to $Lx$).
From \eqref{Eproper} we find that a test string freely falling a proper distance $L$ gains proper energy
\beq\label{EchangeGR}
\Delta \tE_{grav} = \frac{E}{r^z}(e^z-1) \sim \frac{E}{r^z} = \gamma m
\eeq
from the gravitational potential in the Lifshitz geometry.

We would like to compare the proper energy gained from the gravitational potential, $\Delta \tE_{grav}$, with the proper energy lost due to scattering with D0-branes, $\Delta \tE_{scatt}$.  In order to do this, we need to understand the amplitude and cross section of string-D0 scattering.

\subsection{Amplitude and cross section}

We will need to make use of some basic features of closed strings scattering off of D0-branes in the Regge limit. Some of these basic features can be understood in a simple way along the lines of \cite{Lenny}.   Detailed results 
on such amplitudes for the leading Regge trajectory were derived in the works \cite{Monni, veneziano},
to which we refer the reader for further details.  This analysis was done in ten-dimensional flat spacetime, whereas our D0-branes collectively source a Lifshitz geometry.  
As discussed above (\ref{numdenten}), for simplicity we can consider a regime in which the D0-branes are distributed in the ten dimensions in such a way as to be separated by more than string scale.  This, and the fact that the amplitudes we consider have significant support at small (sub-$\sqrt{\alpha'}$) impact parameter, justifies our use of flat space results.   
 
We consider an initial string state with mass and momentum $\alpha' m^2=-\alpha' \tp^2=4n$, which scatters off a stationary D0-brane into a final string state with $\alpha' m'^2=-\alpha' \tp'^2=4n'$.
The proper momentum transfer between the string and D0-brane is $\tq=\tp'-\tp$.
We will make use of the kinematic variables $s=\tE^2\equiv (\tp^0)^2$ and $t=-\tq^2$.  Note that $s$ is different from the usual Mandelstam variable in a two-body scattering process.

We will focus on strings with $\gamma m\gg 1/\sqrt{\alpha'}$ for two reasons. First, as we explained in \S2, the tidal force and its effect on strings and other probes can only diverge if $\gamma$ does.  Secondly, we are interested in the behavior of the mass independently of $\gamma$, to see if the string's proper energy stays bounded with all its interactions with gravity and the stress energy sources taken into account.     
Therefore we will consider amplitudes in the Regge limit, in which $\alpha' s \to\infty$ with $\alpha' t$ fixed. This is the limit in which the incoming string has a very large energy, but the scattering process has a fixed momentum transfer, and hence a small deflection angle. In particular, we will see that the typical momentum transfer $\tq$ is at most of order the string scale, much smaller than the D0-brane mass $m_0\sim 1/(g_s\sqrt{\alpha'})$, so we can ignore the recoil of the D0-brane.  
We will be interested in the energy $\Delta\tE_{scatt}$ transmitted to the D0-branes over a distance of order $L$.  In each collision, the energy transfer is small, approximately $\vec{\tq}^2/(2m_0)$.  In a distance of order $L$ we build up
\beq\label{Elossone}
\Delta \tE_{scatt}\sim N_{collisions}|_L  \frac{\vec \tq^2}{2m_0} 
\eeq
which will turn out sizeable enough to compete with $\Delta \tE_{grav}$.  To see this,  
we need the scattering cross section to determine $N_{collisions}|_L$ as a function of the model parameters.  

In the regime where free string theory describes single-string states accurately, a string at rest behaves like a random walk of length proportional to its mass (with each step being of string scale), and hence occupies a region of size $R_s\sim \sqrt{m}\alpha'^{3/4}\sim n^{1/4}\sqrt{\alpha'}$.  
We can give an estimate for a lower bound\footnote{There could be some enhancement from summing over the final states.  As we will see the lower bound estimate \eqref{tendcross} already bounds $\gamma$ and $m$ from above sufficiently for our discussions.} on the cross section of order
\beq\label{tendcross}
\sigma \simge g_s^2 (m\gamma)^2 \alpha'^5 \,.
\eeq
This behavior follows from the Regge limit amplitude
${\cal A}$, which is proportional to $(\alpha's)^{\frac{\alpha't}{4}+1}$ (with a relativistic normalization of the states).  The differential cross section is proportional to $|{\cal A}|^2/s$, with typical transverse momentum transfer $\tq=\sqrt{-t}$ of order $1/\sqrt{\alpha' \log(s\alpha')}$.  These features can be understood physically as follows.  As explained in \cite{Lenny}, the growth in $s$ is a consequence of the proliferation of accessible partons as one increases the string energy.  The logarithmic falloff of the typical momentum transfer arises from quantum fluctuations of the string embedding coordinates,  cut off by the finite resolution in the process.  The typical momentum transfer of order $1/\sqrt{\alpha' \log(s\alpha')}$ is manifest in the explicit expression for the string-D0 scattering cross section for $n$ of order 1;  one has a phase space integral over $\tq$ which favors larger $\tq$ combined with the exponential suppression of  $\tq\gg 1/\sqrt{\alpha' \log(s\alpha')}$ from the factor of $(\alpha's)^{\frac{\alpha't}{2}}$ in the squared amplitude.  
This string scale momentum transfer (up to the logarithm) should apply also to the scattering of the D0-brane off of a local region of a longer, large-$n$ string; the longer-range interactions between the D0-brane and distant parts of the string are quite weak.  The precise calculation of the amplitude and cross section is very complicated for a typical string state at high level $n$.  
In the appendix, we describe some related aspects of detailed amplitudes in the somewhat simpler case of the leading Regge trajectory \cite{Monni}\ and outline the calculation for more generic states.

Finally, we can use \eqref{scatteringnumber} and \eqref{tendcross} to estimate the energy lost to scattering with D0-branes in a proper distance $L$ as
\beq\label{Echangescatt}
\Delta \tE_{scatt}\sim N_{collisions}|_L  \frac{\vec \tq^2}{2m_0} \simge \frac{g_s^2 m^2\gamma^2\alpha'}{ L} \,,
\eeq
where we have used the typical momentum transfer $\tq\sim 1/\sqrt{\alpha'}$ (again dropping logarithmic terms).
This leads to a ratio
\beq\label{Eratio}
\frac{\Delta \tE_{scatt}}{\Delta \tE_{grav}} \simge \frac{g_s^2 m \gamma\alpha'}{L} \,.
\eeq
These results hold in the regime where the string is much smaller than $L$ since they are based on ten-dimensional flat spacetime scattering theory; this is valid for a range of $N_2$ and $N_6$ consistent with small $g_s$ and large $L$.  As we will note below, this also helps maintain perturbative control for large string mass \cite{GaryJoe}.  In other regimes of parameters, the system may evolve to become effectively four-dimensional, bringing in new effects which would be interesting to analyze in future work.  

\subsection{Asymptotic speed and mass bounds}

We can now put together our results to bound the  late-time velocity and mode production on a test string in the Lifshitz system.  Our bounds will be conservative, since there are other effects we did not include which can also drain energy from the system.  These include processes in which open strings are created on the D0-branes and processes in which the massive strings decay into lighter strings plus gravitational radiation.  

That being said, let us proceed with our conservative bounds.   By the time the ratio (\ref{Eratio}) is of order one or greater, the string loses more proper energy to the D0-branes than it gains by falling down the gravitational potential. 
Let us focus on the would-be singular regime in which tidal forces build up a string mass of order $\gamma/L$ (\ref{tidalmass}).     
The scattering energy loss becomes competitive with the gain from gravity when 
\beq\label{gammalate}
\gamma \sim \frac{\sqrt{L/\alpha'}}{g_s} \,.
\eeq
It occurs at a scale well within the regime of validity of perturbative string theory \cite{GaryJoe, correspondence}.       

The string can lose proper energy by a combination of slowing down or transitioning to lower oscillator number.  The latter clearly counteracts the would-be singularity from the tidal force described in \S2.  In the former case the string slows down, with a late-time subluminal velocity bound of order (\ref{gammalate}).  We found in \S2 that this implies that the tidal force remains finite (\ref{finiteforce}).  In particular, the tidal force contribution to the worldsheet mass squared becomes constant if the string approaches a terminal velocity, shutting off the mode production.        

Before concluding our analysis, however, let us consider whether cumulative effects of the finite tidal force (if it varies with time) -- or the scattering process itself -- can lead to unbounded mode production over the entire trajectory of the string, which we have learned takes infinite proper time to reach the would-be singularity at $r=0$.  A buildup of modes which generated an unbounded string mass $m\sim \sqrt{n/\alpha'}$ could ultimately backreact on the putative Lifshitz solution.  In the regime where the process is effectively ten dimensional, as the mass increases the string becomes more dilute and remains under perturbative control \cite{GaryJoe}.  However, if it were to grow larger than $L$, the physics would be come effectively four dimensional, potentially leading to stronger self-interactions invalidating a perturbative analysis.\footnote{We say ``potentially" here because the self-gravitation of the four dimensional strings in Lifshitz will be affected by its curvature, necessitating a generalization of the analysis in \cite{GaryJoe}.}    

We can in fact put an upper bound on the asymptotic value of $m$ (or equivalently $n$) within the ten-dimensional regime.  The only way that a test string can maintain a large mass $m$ while traveling in this Lifshitz background is to somehow maintain $\Delta\tE_{scatt} \simle \Delta\tE_{grav}$ (at least asymptotically) -- if not, the proper energy of the string would be continuously drained until the remaining energy becomes smaller than the initial $m$ (no matter how large $\gamma$ started to be), meaning that the string would not be able to maintain its initial mass.  
Requiring the ratio \eqref{Eratio} to be no larger than 1, we obtain 
\beq\label{mbound}
m \simle \frac{L}{g_s^2\gamma\alpha'}\le  \frac{L}{g_s^2\alpha'}
\eeq
which we interpret as an upper bound on the asymptotic value of $m$.  In order to maintain our ten-dimensional description, we must impose that $m<L^2/\alpha'^{3/2}$ so that the string fits within the internal $S^6$.  A sufficient condition for this is $L>\sqrt{\alpha'}/g_s^2$, and it is viable in our model.\footnote{Away from this regime, it would be interesting to study the implications of strings which do collapse into black holes, but this is beyond the scope of the present work.}      

Altogether, the full system -- including the stress-energy sources intrinsic to the solution -- has a built-in feedback mechanism to prevent divergent mode production.\footnote{Some stress is healthy, helping to avoid certain dangers.}  
 The two basic effects of the metric -- acceleration toward the speed of light and oscillator excitations due to tidal forces -- would be very dangerous by themselves.  However, from (\ref{Echangescatt}) we see that they enhance $\Delta \tE_{scatt}$, self-consistently preventing either effect from growing out of control.  The string does not approach the speed of light, and thus as explained in \S2\ the tidal forces acting on it remain finite throughout its evolution.  Its cumulative mode production from all effects is also bounded as just explained.  The string slows down, propagates forever, and never gets too excited.

\section{Conclusion}\label{conclusion}

We have found that in UV complete Lifshitz spacetimes, stress-energy sources can modify propagation strongly enough to avoid singular tidal forces.  In particular, in the examples \cite{HPST}\ a constant proper D0-brane density functions as a scattering target which slows down strings sufficiently to avoid divergent tidal forces.  Probe D-branes also interact with the D0 density, draining energy into stretched open strings \cite{trapping}.   It would be very interesting to apply the same idea to other UV completions of Lifshitz spacetimes, such as \cite{Balas, Donos, Gregory, Cassani:2011sv}\ and to examples such as \cite{Edgar}.  In those cases, we will need to analyze the interaction of strings and other probes with a constant proper energy density contained in various fluxes supporting the geometry.   
It would also be interesting to see how the effects considered here can be applied to more general non-vacuum solutions, and investigate whether they can resolve other types of singularities.

\bigskip
\centerline{\bf{Acknowledgements}}
We would like to thank D. Harlow, S. Hartnoll, G. Horowitz, S. Kachru, K. Narayan, J. Polchinski, S. Shenker, and G. Torroba for useful discussions.
We thank the Kavli Institute for Theoretical Physics for hospitality during parts of this project.  This work is supported in part by the National Science Foundation 
under grant PHY05-51164, by the NSF under grant PHY-0756174, and by the Department of Energy under contract DE-AC03-76SF00515.
SH is supported by the ARCS 
Foundation, Inc.\ Stanford Graduate Fellowship.

\appendix 

\section{Details on the amplitude and cross section}
In this appendix we show the consistency of the estimate \eqref{tendcross} on the Regge limit cross section with known analytic results from string perturbation theory on the scattering amplitude of a closed string with a D0-brane \cite{Monni}.     

For simplicity we will focus on string states on the leading Regge trajectory, for which the scattering amplitude is given by\footnote{The scattering amplitude in \cite{Monni} is based on non-relativistic normalization for the D-brane in/out states (and the usual relativistic normalization for the string states).  We use the relativistic normalization for all states in \eqref{Ann}, and therefore differ from the amplitude in \cite{Monni} by $\sqrt{2m_02(m_0^2+\vec\tq^2)^{1/2}}\approx 2m_0$.}
\beq\label{Ann}
A_{n,n'}={\hat{\cal N}} \alpha'^3 K_{n,n'}(\tq,\epsilon,G)e^{-i\pi\alpha' t/4}\Gamma\left(-\frac{\alpha't}{4}\right)(\alpha's)^{\frac{\alpha't}{4}+1} \,,
\eeq
where ${\hat{\cal N}}$ is a normalization constant of order 1, and $K_{n,n'}$ is a kinematic function
\begin{multline}\label{Knn}
K_{n,n'}(\tq,\epsilon,G)=\frac{1}{n!n'!} \left(\frac{\alpha'}{2}\right)^{n+n'} \sum_{a,b=0}^{\min \{n,n'\}} \left(-\frac{\alpha'}{2}\right)^{-a-b} C_{n,n'}(a) C_{n,n'}(b) \\
\times \tq^{n'-a} \cdot \mc G_a \cdot \varepsilon_a \cdot \tq^{n-a} \, \tq^{n-b} \cdot \tilde\varepsilon_b \cdot \tilde{\mc G}_b \cdot \tq^{n'-b}
\end{multline}
which depends on the momentum transfer $\tq$ and the polarization tensors $\epsilon$, $G$ of the initial and final string states.  Here the combinatoric factor $C_{n,n'}$ is defined as
\beq
C_{n,n'}(a) \equiv \frac{n! n'!}{a! (n-a)! (n'-a)!} \,.
\eeq
Note that $\varepsilon$, $\tilde\varepsilon$ (or $\mc G$, $\tilde{\mc G}$) are holomorphic and anti-holomorphic components of $\epsilon$ (or $G$) respectively.  They are totally symmetric, transverse, and traceless, due to physical state conditions required by BRST invariance.  The term $\tq^{n'-a} \cdot \mc G_a \cdot \varepsilon_a \cdot \tq^{n-a}$ is defined by contracting $a$ indices of $\mc G$ with $\varepsilon$, and their remaining indices with powers of $\tq$.

The sum \eqref{Knn} comes from different contractions in the disc amplitude with two closed string vertex operators.  It is in general difficult to estimate, as it contains terms with alternating signs.  For simplicity we will set $n'=n$.  This provides a lower bound estimate on the total cross section.  To simplify our calculation even further, let us choose the initial momentum $\vec\tp$ to be in the 9-direction, and choose an initial polarization such that the only nonzero components of $\varepsilon$ are $\varepsilon_{11\cdots1}=-\varepsilon_{22\cdots2}=1/\sqrt{2}$ (and similarly for $\tilde\varepsilon$).  This is indeed transverse to $\tp$.

The final polarization must be transverse to $\tp'=\tp+\tq$, and we choose one defined by $\mc G_{1'1'\cdots1'}=-\mc G_{2'2'\cdots2'}=1/\sqrt{2}$ in the rest frame of the final string (and similarly for $\tilde{\mc G}$).  Again this provides a lower bound on the cross section.  We can find the components of $\mc G$ in the ``lab'' frame (where the initial D0-brane is at rest) by making a Lorentz boost in the $\vec\tp'$ direction.  As we will see in a moment, this boost is almost parallel to $\vec\tp$ because the typical momentum transfer $\vec\tq$ is much smaller than $\vec\tp$, and therefore $\mc G$ is approximately equal to $\varepsilon$.  This can be verified by noting that the boost from the rest frame has $\partial X'^i/\partial X^j \approx \delta_j^i +\mc O(\frac{\tq^2}{\gamma m^2})$, and therefore the spatial components of $(\mc G-\varepsilon)$ are at most of order $\big([1+\mc O(\frac{\tq^2}{\gamma m^2})]^n-1\big)\sim \alpha'\tq^2/\gamma$, which is small for $\tq$ smaller than the string scale.  Therefore in our estimate below we set $\mc G\approx \varepsilon$.

One way of estimating the alternating sum $\eqref{Knn}$ is to consider a regime where it is dominated by a single term.  Let us focus on the sum over $a$.  The $a=n$ term is of order 1, and the $a=n-1$ term is of order $\alpha'n(\tq_1^2+\tq_2^2)$. Therefore for $\tq_1,\tq_2\simle 1/\sqrt{n\alpha'}$, the $a=n$ term is dominant and the net sum $K_{n,n}$ is of order 1.  This provides a conservative lower bound on the cross section, which is given by
\beq\label{cross}
\sigma \sim \int\frac{d\tq_1 d\tq_2 \cdots d\tq_8}{\tE^2 m_0^2}|A_{n,n}|^2 \,.
\eeq
The $\tq_9$ component is fixed in terms of the other eight spatial components (collectively called $\tq_\perp$) to be of order $\tq_\perp^2/|\vec \tp|$.  It is much smaller than $\tq_\perp$ because $\tq_\perp$ is less than string scale (as we will argue in a moment) while $\vec \tp$ is much larger than string scale.

The factor $(\alpha's)^{\alpha't/4}$ in \eqref{Ann} imposes that the typical $\tq=\sqrt{-t}$ be of order $1/\sqrt{\alpha'\log(\alpha's)}$, smaller than the string scale when $\alpha's$ is large.  At this typical $\tq$, the amplitude $A_{n,n}$ is of order $\alpha'^4 K_{n,n} s$, where we have dropped powers of $\log(\alpha's)$.  Using our conservative estimate that $K_{n,n}$ is of order 1 when $\tq_1,\tq_2\simle 1/\sqrt{n\alpha'}$, we find that the cross section \eqref{cross} is at least of order
\beq
\sigma \simge \alpha'^4 g_s^2 \gamma^2 \,,
\eeq
where we have used $s=\tE^2=(\gamma m)^2\sim n\gamma^2/\alpha'$ and $m_0\sim 1/(g_s\sqrt{\alpha'})$.

This result is almost the same as the generic estimate \eqref{tendcross}, differing in the small values of $\tq_1$ and $\tq_2$.  This is an artifact of the very special state we have chosen: the string is at a very high oscillator level in the 1- and 2-directions, well described as classically stretching and contracting in these directions.  Locally on the string, there is approximate translational invariance in the 1- and 2-directions, so the D0-brane that impinges on it cannot absorb much momentum in these directions (until it detects the finite length of the string).  For generic states, there will be no such symmetry, and the D0-brane should be able to absorb momentum up to order $1/\sqrt{\alpha'\log(\alpha's)}$ in all directions (perpendicular to $\vec\tp$).  In particular, a small-$n$ string can impart such an isotropic momentum transfer, and in general this should also hold for a longer string made up of many bits.  

Indeed a generic string state at large $n$ would have excitations in higher modes $\alpha_{-k}^\mu$, instead of having only $\alpha_{-1}^\mu$ as is the case for the leading Regge trajectory.   It is difficult to calculate the amplitude and cross section explicitly for more generic string states, but we can explore their structure as follows. Consider for example schematically a vertex operator whose bosonic part is $\partial X^{\mu_1} \partial^2 X^{\mu_2} \cdots \partial^r X^{\mu_r} \times \text{(anti-holomorphic part)}\times e^{i\tp\cdot X}$.  Here $r$ is chosen to be of order $\sqrt{n}$ so that the total oscillator level is equal to $n$.  The number of states of this kind is exponential in $\sqrt{n}$, similar to the Hagedorn density of states, so this class of vertex operators describes much more  generic states than the leading Regge trajectory.  

Assuming this set of vertex operators includes examples satisfying the physical state conditions, the scattering amplitude is calculated by summing over all possible mutual contractions.  We would like the dominant contribution to come from full contractions of the $\partial^k X$ factors among themselves.  This is the analog of the $a=n$ term in \eqref{Knn}.  The subleading contribution to the amplitude comes from contractions in which a previously contracted pair of $\partial^k X$ breaks up and is contracted with $e^{i\tp\cdot X}$ (and $e^{i\tp'\cdot X}$) instead.  There are many such pairs to choose from, and this is the why the $a=n-1$ term in \eqref{Knn} is enhanced by a factor $n$.  However, here a pair such as $\partial^{k_1} X \partial^{k_2} X$ does not want to break up if $k_1$ and $k_2$ are both large, because their contraction gives a factor of $(k_1+k_2-1)!$ whereas being contracted with $e^{i\tp\cdot X}$ only gives $(k_1-1)! (k_2-1)!$, which is exponentially smaller than $(k_1+k_2-1)!$ at large $k_1$ and $k_2$.  Therefore only pairs with a small $k_1$ or $k_2$ can break up and contribute significantly.  There are $\mc O(1)$ of these, and therefore the subleading contribution to the amplitude is smaller than the dominant contribution by a factor of order $\tq^2$.  Hence $\tq$ can be of order $1/\sqrt{\alpha'\log(\alpha's)}$ while keeping the amplitude of order $\alpha'^4 s$.  This gives $\sigma\sim \alpha'^4 g_s^2 n \gamma^2$ which agrees with \eqref{tendcross}.

\bibliographystyle{JHEP}
\renewcommand{\refname}{Bibliography}
\addcontentsline{toc}{section}{Bibliography}
\providecommand{\href}[2]{#2}\begingroup\raggedright

\end{document}